\documentclass[12pt]{iopart}

\usepackage{graphicx}

\begin{document}
\title[How to confirm and exclude different models in the Casimir effect]{How to
confirm and exclude different models of material properties in the Casimir effect}

\author{ V~M~Mostepanenko
}

\address{Central Astronomical Observatory
at Pulkovo of the Russian Academy of Sciences,
St.Petersburg, 196140, Russia}
\address{Institute of Physics, Nanotechnology and
Telecommunications, St.Petersburg State
Polytechnical University, St.Petersburg, 195251, Russia}

\ead{vmostepa@gmail.com}

\begin{abstract}
We formulate a method allowing to confirm or exclude the
alternative
models of material properties at some definite confidence level
in experiments on measuring the Casimir force. The method is
based on the consideration of differences between the theoretical
and mean measured quantities and the confidence intervals for
these differences found at sufficiently high or low confidence
probabilities. The developed method is applied to the data of
four recent experiments on measuring the gradient of the
Casimir force by means of a dynamic atomic force microscope.
It is shown that in experiments with Au-Au and Ni-Ni test
bodies, where the Drude model approach is excluded at a 95\%
confidence level, the plasma model approach agrees with
the data at
higher than 90\% confidence. In experiments using an Au
sphere interacting with either a Ni plate or a graphene-coated
substrate the measurement data agree with the common prediction of
the Drude and plasma model approaches and theory using the
polarization tensor at 90\% and 80\% confidence levels,
respectively.
\end{abstract}


\maketitle

\section{Introduction}

The Casimir effect \cite{1} manifests itself as a force acting
between closely spaced material bodies. Similar to the
van der Waals force, the Casimir force is caused by the zero-point
and thermal fluctuations of the electromagnetic field \cite{2}.
The only difference is that the Casimir force acts at larger
separation distances and depends on the relativistic retardation.
The Casimir effect is a multidisciplinary phenomenon important
for condensed matter physics, nanotechnology, elementary particle
physics, atomic physics, and for gravitation and cosmology (see
the monographs [3--5] and references therein).
Recently a lot of experiments on measuring the Casimir force
between metallic, semiconductor and dielectric test bodies have
been performed (see reviews [6--8] and more recent results
[9--23]). The experimental data have been compared with
theoretical predictions of the fundamental Lifshitz theory
\cite{24} expressing the Casimir force in terms of the
frequency-dependent dielectric permittivities of material
bodies.

The theory-experiment comparison revealed a fundamental puzzle
in Casimir physics. It turned out that for metallic test bodies
the theoretical results are in agreement with the measurement
data only under a condition that the relaxation properties of
conduction electrons are not taken into account in computations
[3,6,15,17,25--27]. Technically this means that in calculation
of the thermal Casimir force an extrapolation of the measured optical
data for the complex index of refraction to zero frequency
should be made using the nondissipative plasma model rather
than the Drude model taking the relaxation properties of
conduction electrons into account. There is one experiment \cite{12}
claiming an agreement of the data with the Drude model
extrapolation. This experiment, however, measured not the
Casimir force, but up to an order of magnitude larger force
supposedly originating from the electrostatic patches.
The latter force was modeled using the two fitting parameters
and subtracted from the data (critical analysis of such a
procedure can be found in the literature [29--30]).
Another facet of the puzzle is that for the test bodies made
of dielectric materials the measured
thermal Casimir force is in
agreement with the predictions of the Lifshitz theory only
if the contribution of free charge carriers to the dielectric
permittivity is omitted [3,6,7,11,13,31--34].
It was also shown [3,6,7,35--38] that the Lifshitz theory
taking into account the relaxation properties of conduction
electrons in metals with perfect crystal lattices or free
charge carriers in dielectrics violate the third law of
thermodynamics (the Nernst heat theorem).

Attempts to resolve the puzzle in Casimir physics have
raised a question on how to make the comparison between
experiment and theory more rigorous.
At separations below a few hundred nanometers, where
measurements of the Casimir force are most precise,
the thermal effect predicted by the Drude model does not
exceed a few percent of the measured force. As to the
thermal effect predicted by the plasma model, at short
separation distanves it is below the instrumental
sensitivity for all materials with exception of
graphene \cite{40}.
Because of this, up to the present the main attention was
paid to statistical procedures allowing to exclude some model
of material properties at sufficiently high confidence level.
As was already mentioned, many experiments on the
thermal Casimir
force resulted in an exclusion of the Drude model for metals
and of the role of conduction electrons for dielectrics.
In so doing it was sometimes assumed that
 the confidence levels for the exclusion of the
Drude model by the data and for the agreement of the same
data with the plasma model are common \cite{25}.

In this paper, we discuss how to confirm or exclude some
alternative models of material properties when comparing
measurements of the thermal Casimir force with theory.
We demonstrate that if there are two competing models
of material properties, the confidence levels for the
exclusion of one of them and for the confirmation of
another one are usually different.
For this purpose we consider the random quantity
$F_{\rm diff}^{\prime}(a)$ equal to differences between
theoretical and mean experimental values of the measured
gradient of the Casimir force as functions of separation
$a$ between the test bodies. In order to exclude some model
of material properties, we calculate the confidence
intervals  $[-\Xi(a),\Xi(a)]$ for this quantity at
sufficiently high confidence level. Then we verify,
whether the values of $F_{\rm diff}^{\prime}(a)$
belong to it. This approach was often used in previous
literature [3,6,7,18,25,27,31,32].
To make sure that some model of material properties
is in agreement with the data, we
determine  the confidence interfals for
$F_{\rm diff}^{\prime}(a)$ at sufficiently low confidence
level. If it is found that the model is not excluded
even at this low confidence level, one can conclude that
it agrees with the data at a high confidence level complimentary
to unity. This approach was prefiously described only
briefly and only for a measurement of the Casimir force
between two Ni test bodies \cite{18}. Here we apply it
to measurements of the gradient of the Casimir force
between two Au test bodies \cite{15}, between one Au
and one Ni test bodies \cite{14}, and between an Au test
body and a graphene-coated substrate \cite{20}.
In all these cases we determine the confidence levels
with which the respective theory is in agreement with the
measurement data.

The paper is organized as follows. In section 2 we
introduce the main notations and briefly formulate the
results for Ni-Ni test bodies. In section 3 the case of
a Au-coated sphere interacting with a Au-coated plate is
considered. Section 4 is devoted to the experiment with
a Au-coated sphere interacting with a Ni-coated plate.
In section 5 a Au-coated sphere interacting with a
graphene-coated substrate is considered. In section 6
the reader will find our conclusions and discussion.

\section{Selection between models of material properties in
Casimir experiment with two Ni test bodies}

In all experiments under consideration in this paper,
the gradient of the thermal Casimir force
$F^{\prime}(a)$, acting between the sphere and the plate
spaced at a separation $a$,
is measured by means of a dymanic atomic
force microscope (AFM). There is a standard method to
represent the measurement data as crosses centered at
points with coordinates
$[a_i,\bar{F}_{\rm expt}^{\prime}(a_i)]$, where
$\bar{F}_{\rm expt}^{\prime}(a_i)$ is the mean gradient
of the Casimir force measured at a separation $a_i$,
the length of the horizontal arms is equal to
$2\Delta^{\!\rm tot} a_i$ and the height of the vertical arms is equal
to $2\Delta^{\!\rm tot}\bar{F}_{\rm expt}^{\prime}(a_i)$
\cite{3}. Here, the total errors (i.e., the random and systematic
combined), $\Delta^{\!\rm tot} a_i$ and
$\Delta^{\!\rm tot}\bar{F}_{\rm expt}^{\prime}(a_i)$, are meant
to be determined a some common confidence level.
If the theoretical band does not overlap with the majority of
crosses within some range of separations, it is said that this
theory (theoretical model) is excluded by the data at a given
confidence level. In doing so, the width of the theoretical band
is equal to twice the theoretical error,
$\Delta^{\!\rm tot}F_{\rm theor}^{\prime}(a)$ with which
theoretical values of the force gradient  are computed
\cite{3,6}.  Although this method of experiment-theory
comparison allows to determine the confidence level of exclusion
of some model by the data, it is somewhat difficult to
quantitatively
characterize the measure of agreement when there is a partial
overlap between the experimental crosses and the theoretical band.

Another method to compare experiment with theory is based on
the consideration of the random quantity
\begin{equation}
F_{\rm diff}^{\prime}(a)=F_{\rm theor}^{\prime}(a)-
\bar{F}_{\rm expt}^{\prime}(a).
\label{eq1}
\end{equation}
\noindent
Here, the theoretical force gradients are computed with the help
of the Lifshitz theory using some model of dielectric response
and taking into account the surface roughness.
The confidence interval for this quantity at a given confidence
level $\beta$ can be expressed via the total experimental and
theoretical errors. The conservative estimation for the halfwidth
of this interval at a separation $a_i$ is given by \cite{3,6}
\begin{eqnarray}
&&
\Xi_{F_{\rm diff}^{\prime}}^{\beta}(a_i)=\min\left\{
\Delta^{\!\rm tot}F_{\rm theor}^{\prime}(a_i)+
\Delta^{\!\rm tot}\bar{F}_{\rm expt}^{\prime}(a_i),
\vphantom{\sqrt{\left[\Delta^{\!\rm tot}F_{\rm theor}^{\prime}(a_i)
\right]^2}}\right.
\nonumber \\
&&
~~~~~~~~~~~~~~~~~~~~~~~~~~~~~~
\left.
k_{\beta}\sqrt{\left[\Delta^{\!\rm tot}F_{\rm theor}^{\prime}(a_i)
\right]^2+
\left[\Delta^{\!\rm tot}\bar{F}_{\rm expt}^{\prime}(a_i)\right]^2}
\right\}.
\label{eq2}
\end{eqnarray}
\noindent
Here, $k_{\beta}$ is a tabulated coefficient from the composition
law of two uniform distributions \cite{41}. As an example, for
$\beta=0.95$ (a 95\% confidence level) $k_{\beta}=1.1$.
If for some theoretical model of material properties more than
$1-\beta$ percent of the
differences $F_{\rm diff}^{\prime}$ are outside the confidence
interval
\begin{equation}
[-\Xi_{F_{\rm diff}^{\prime}}^{\beta}(a_i),
\Xi_{F_{\rm diff}^{\prime}}^{\beta}(a_i)]
\label{eq3}
\end{equation}
\noindent
for all $a_i$ belonging to some interval $[a_{\min},a_{\max}]$,
one can conclude that this model is experimentally {\it excluded}
for separations from $a_{\min}$ to $a_{\max}$ at a confidence
level $\beta$. Alternatively, a model of material properties, for
which no fewer than $\beta$ percent of the differences
$F_{\rm diff}^{\prime}$ belong to the confidence interval
(\ref{eq3}) within any separation subinterval from
$a_{\min}$ to $a_{\max}$, is {\it consistent} with the
measurement data within this confidence interval.
In fact, this comparison method is close in spirit to the
so-called ``p-value" in statistics \cite{39a}, except in our case
it is applied over the aggregate of the measurements done at
different separations. Note, however, that the method of p-value
is most often applied for the verification of the null hypothesis
in psychological, medical, and economy research which are more
empirical than physical research. The latter is often based on the
well-established fundamental theory.

Let us assume that some theoretical model is {\it consistent} with the
data within the confidence interval defined
at some high confidence level $\beta$. It should be
particularly
emphasized that this does not mean that the model is
{\it confirmed} by the data, or, synonymously, agrees with
the data at the same high confidence.
Really, the higher is the confidence propability $\beta$, the
wider is the confidence interval (\ref{eq3}).
For example, if some theoretical approach is excluded at a 95\%
confidence level, this means that the model agrees with the data at only
5\% probability.
At this point, a terminological note is pertinent. In fact it is common to
speak about confirmation of physical theories by the measurement data.
It should be remembered, however, that this ``confirmation" is not
absolute, so that some theory can be confirmed at rather high
confidence level $\beta_1$ (i.e., not excluded at a confidence level
$1-\beta_1$), but excluded at a confidence level $\beta_2<1-\beta_1$.
Keeping this in mind and to avoid confusion, we will subsequently
speak about ``agreement of the model with the data at some
confidence level" as a synonym for ``confirmation of the model by
the data at some confidence level".

Let us consider in more detail the difference
between the concepts of {\it consistency} and {\it confirmation}.
If almost 100\% of the differences (\ref{eq1}) belong to the
confidence interval (\ref{eq3}) defined at a confidence level
$\beta=0.95$ (i.e., the model under consideration is
{\it consistent} with the data within the 95\% confidence
interval),
this does not mean a high degree of {\it agreement}.
Just on the contrary, belonging of the differences (\ref{eq1})
to a sufficiently wide interval (\ref{eq3}) may mean rather poor
agreement between the theoretical and experimental force
gradients. Specifically, some theoretical model consistent with
the data within a 95\% confidence interval might be excluded by
the same data at a 67\% confidence level.
It can be concluded that the degree of {\it agreement} of
some theoretical model increases {\it if a sufficient fraction of
the differences {\rm (\ref{eq1})} is found inside the confidence
interval {\rm (\ref{eq3})} defined at rather low confidence level}
(e.g., $\beta=0.1$).
For instance, let one finds more than 10\% of
the differences (\ref{eq1}) inside such a confidence
interval (\ref{eq3}) within any separation subinterval.
This would mean that the theoretical model is not excluded by
the measurement data even at a 10\% confidence level, or,
equivalently, that this model is in agreement with the data at a higher
than 90\% confidence level.

To make the above formulations more accessible, we illustrate
them by the data of experiment on measuring the gradient of the
Casimir force between a Ni-coated sphere of $R=61.71\,\mu$m
radius and a Ni-coated plate by means of dynamic AFM \cite{17,18}.
The mean measurement data for the gradient of the Casimir force
at $T=300\,$K, $\bar{F}_{\rm expt}^{\prime}(a_i)$, were compared
with the theoretical predictions of the Lifshitz theory
\cite{3,24,42} describing the dielectric properties of Ni by means
of the frequency-dependent dielectric permittivity
$\varepsilon^{\rm Ni}(\omega)$.
The magnetic properties of Ni were described by the
Debye formula \cite{18} for the frequency-dependent magnetic
permeability $\mu(\omega)$. For ferromagnetic materials $\mu(i\xi)$
becomes equal to unity at $\xi>10^{5}\,$Hz, i.e., starting from
frequencies much smaller than the first Matsubara frequency at
room temperature. From this it follows that the magnetic Casimir
interaction is determined by the static magnetic permeability
$\mu^{\rm Ni}(0)=110$ (quick decrease of $\mu(i\xi)$ to unity
with increasing $\xi$ and its role in the magnetic Casimir effect
was discussed in \cite{43}).

The dielectric properties of Ni were described using two
different models. Within the first model, the dielectric
permittivity at the imaginary Matsubara frequencies was obtained
using the standard Kramers-Kronig relation from
\begin{equation}
{\rm Im}\,\varepsilon(\omega)=2n_1(\omega)n_2(\omega),
\label{eq4}
\end{equation}
\noindent
where
$n_1(\omega)$ and $n_2(\omega)$ are the real and imaginary parts
of the complex index of refraction, respectively, measured over
a wide frequency region \cite{44}. Below the minimum frequency,
where the optical data are not available,
${\rm Im}\,\varepsilon(\omega)$ was extrapolated down to zero
frequency by means of the Drude model (the so-called
{\it Drude model approach})
\begin{equation}
\varepsilon_D(\omega)=1-\frac{\omega_p^2}{\omega[\omega+
i\gamma(T)]}.
\label{eq5}
\end{equation}
\noindent
Here, for Ni the plasma frequency $\omega_p=4.89\,$eV and the
relaxation parameter at room temperature
$\gamma(T)=0.0436\,$eV \cite{44,45} were used.

Within the second model, the optical data with subtracted
contribution
of conduction electrons are used to find the parameters of
oscillators describing the bound (core) electrons. Then the
dielectric permittivity is represented in the form
\begin{equation}
\varepsilon_{gp}(\omega)=1-\frac{\omega_p^2}{\omega^2}+
\sum_{j=1}^{K}\frac{g_j}{\omega_j^2-\omega^2-
i\gamma_j\omega},
\label{eq6}
\end{equation}
\noindent
where $K$ is the number of oscillators, $g_j$ are the
oscillator strengths, and $\gamma_j$ are the relaxation
frequencies (this is the so-called {\it generalized
plasma-like model}). Thus, in the second model of
dielectric properties the optical data should be rid off
the contribution of free electrons and extrapolated to
zero frequency by means of the nondissipative plasma
model
\begin{equation}
\varepsilon_p(\omega)=1-\frac{\omega_p^2}{\omega^2},
\label{eq7}
\end{equation}
\noindent
describing the plasma of free electrons (the so-called
{\it plasma model approach}).
Equivalently, the same results can be obtained directly
from the imaginary part of the dielectric permittivity
(\ref{eq4}), where the optical data are rid off the
contribution of free electrons, by using the generalized
Kramers-Kronig relation applicable to dielectric functions
having the second-order pole at zero frequency
\cite{3,6,47}.

It is necessary to stress that at low (quasistatic)
frequencies  the Maxwell equations lead to the dielectric
permittivity of the Drude-type (\ref{eq5}) which is inverse
proportional to the first power of frequency \cite{46}.
As to the plasma model (\ref{eq7}), it is applicable  in the
region of high (infrared) frequencies satisfying the
condition $\omega\gg\gamma(T)$. Because of this, an exclusion
of the Drude model approach and consistency with the plasma
model approach in the most of experiments on measuring the
Casimir force discussed in section 1 is considered as a puzzle.

Now let us return to the experiment \cite{17,18} dealing with
two Ni test bodies and consider the quantitative measures of
exclusion and agreement for different models of material
preperties.
In figure 1(a,b) we present as dots the differences of the
gradients of the Casimir force (\ref{eq1}) as functions of
separation computed using the Drude model (the upper set of
gray dots) and the plasma model (the lower set of black dots)
approaches. The upper and lower solid lines indicate the
borders of the confidence intervals (\ref{eq3}) as functions
of separation computed using (\ref{eq2}) at (a) 95\% and
(b) 67\% confidence levels. Note that the borders of the
confidence intervals very slightly depend on the theoretical
approach used and are mostly determined by the chosen
confidence level. As is seen in figure 1(a,b), the Drude
model approach is excluded by the data over the separation
region from 223 to 345\,nm at a 95\% confidence level and
from 223 to 420\,nm at a 67\% confidence level, respectively.
Here, not a bit more than 5\% (respectively, 33\%) of all
dots are outside the confidence intervals, as required for
the exclusion of the model, but almost 100\% do not belong
to the confidence intervals. These are rather strong evidences
to believe that this approach is not adequate.

{}From figure 1(a,b) it is also seen that the plasma model
approach is consistent with data at both 95\% and 67\%
confidence levels  over the entire measurement range from
223 to 550\,nm. As explained above, however, the consistency
at rather high confidences (i.e., for rather wide confidence
intervals) does not automatically mean that the approach
agrees with the data at a high probability. To find the measure of
agreement between the plasma model approach and the data,
it is necessary to consider sufficiently low confidence
probability $\beta$ (i.e., narrow confidence interval)
such that at slightly smaller $\beta$ this approach is
already excluded by the data. In figure 2 we plot as black
dots the same differences (\ref{eq1}) computed using the
plasma model approach as are shown in figure 1(a,b).
The upper and lower solid lines indicate the borders of
the confidence intervals found at a 10\% confidence level.
As is seen in figure 2, more than 10\% of all dots within
any separation subinterval are inside the 10\% confidence
intervals. This allows one to conclude that the plasma
model approach is not excluded by the data even at
sufficiently low 10\% confidence level, i.e., this
approach agrees with the data at no less than
90\% confidence.

\section{Selection between models of material properties
in the Casimir experiment with two Au test bodies}

Here, we consider the choice between the plasma and the
Drude model approaches in the experiment on measuring the
gradient of the Casimir force between two Au-coated
surfaces of a sphere of $R=41.3\,\mu$m radius and a plate
by means of dynamic AFM \cite{15}. In this experiment the
Drude model approach was excluded at a 67\% confidence
level over the separation region from 235 to 420\,nm
using a standard method of theory-experiment comparison
described in the beginning of section 2. Later on,
using the statistical method applied here, it was shown
that the data of this experiment exclude the Drude model
approach at a higher 95\% confidence level
over the separation region from 235 to 330\,nm.
The plasma model approach was found to be consistent with
the data at both confidence levels. However, the confidence
level for its agreement with the data was not found.

Before presenting the resolution of this question, it is
pertinent to note that recently a distinction between the
predictions of the Drude and plasma model approaches for two
bodies made of nonmagnetic metal (such as Au) has been
questioned \cite{48}. Specifically, it was claimed \cite{48}
that for a nondissipative plasma model the Lifshitz formula
in the form of summation over the Matsubara frequencies
undergoes a modification. As a result, the discontinuity
between the predictions of the Drude and plasma model
approaches disappears \cite{48}. According to \cite{48},
the physical reason for this is the contribution of the
Foucault currents which plays a role even in the limit
of vanishing dissipation, in contrast to commonly accepted
views.

In this connection we underline that the above claims of
\cite{48} are based on a terminological confusion and do
not contain any new physics. The nondissipative plasma
model is defined in \cite{48} not in accordance to the
conventional definition (\ref{eq7}), but as the limiting
case of the Drude model (\ref{eq5}) when $\gamma\to 0$.
This would be really correct if only nonzero frequencies
$\omega\neq 0$ were considered. However, \cite{48} employs
this definition at all frequencies including
$\omega=0$. Such an employment is mathematically
unjustified because
\begin{eqnarray}
\lim_{\gamma\to 0}\varepsilon_D(\omega)&=&
1-\lim_{\gamma\to 0}\frac{\omega_p^2}{\omega^2+\gamma^2}+
i\frac{\omega_p^2}{\omega}\lim_{\gamma\to 0}
\frac{\gamma}{\omega^2+\gamma^2}
\nonumber\\
&=&
1-\frac{\omega_p^2}{\omega^2}+i\pi\frac{\omega_p^2}{\omega}
\delta(\omega).
\label{eq8}
\end{eqnarray}
\noindent
It is seen that the limiting
case (\ref{eq8}) of the Drude model (\ref{eq5})
when $\gamma\to 0$
coincides with the plasma model (\ref{eq7}) only at
$\omega\neq 0$. At $\omega=0$ it contains an infinitely
large imaginary part, i.e., is not a nondissipative
dielectric permittivity.
Thus, the expression (\ref{eq8}) does not coincide with the
plasma model (\ref{eq7}) and cannot be called ``the plasma
model''. Note also that in accordance to previously
published results [48--50] it is just the contribution of
the Foucault currents which determines a difference in the
theoretical predictions of the Drude and plasma model
approaches.

Now we return to the measurement data of the experiment
\cite{15} measuring the gradient of the Casimir force between
two Au-coated surfaces of a sphere and a plate.
In figure 3(a,b) we present as dots the differences (\ref{eq1})
at different
separations found using the Drude model (the lower sets of
gray dots) and the plasma model (the upper sets of black dots)
approaches. The pairs of solid lines indicate the
borders of the confidence intervals (\ref{eq3}) at different
separations found using (\ref{eq2}) at (a) 95\% and
(b) 67\% confidence levels. The comparison with figure 1(a,b)
shows that for Ni and Au the differences
$F_{\rm diff}^{\prime}$ computed using the Drude model
approach have the opposite signs. This was used \cite{17,18}
to exclude the role of any unaccounted systematic effect in
the measurement data. As was mentioned in the beginning of this
section, the Drude model approach is excluded by the data,
whereas the plasma model approach is experimentally
consistent.

To characterize the measure of agreement between
the plasma model approach and the measurement data, in
figure 4 we plot as black dots the same differences as shown
by the black dots in figure 3(a,b).
The pairs of solid lines in figure 4 indicate the
borders of the confidence intervals found  at a
10\% confidence level. It can be seen that more than 10\% of all
dots within any separation subinterval belong to the
confidence intervals. Thus, the plasma model approach is not
excluded by the data at a 10\% confidence level.
As a result, one can say that in the experiment with two
Au surfaces the plasma model approach agrees with
the data at no less
than 90\% confidence level. This conclusion is the same as was done in
previous section with respect to the experiment with two
Ni surfaces.

\section{Agreement with the models of material properties
in the Casimir experiment with Au-Ni test bodies}

In the experiment \cite{14} the gradient of the Casimir force
was measured between a Au-coated sphere of $R=64.1\,\mu$m
radius and a Ni-coated plate by means of dynamic AFM.
For this configuration within the used separation region
from 220 to 500\,nm the theoretical predictions of the
Drude and plasma model approaches almost coincide.
Thus, the magnitudes of relative differences in the
predictions of both models at separations 220, 300, 400, and
500\,nm are only 0.5\% , 0.2\%, 0.4\%, and 1.2\%,
respectively \cite{14}. This is below the instrumental
sensitivity at respective separations.
At larger separations the relative differences between the
two approaches are much larger. For example, at separations
of 3 and $5\,\mu$m they are equal to 31.1\% and 42.8\%,
respectively \cite{14}. Because of this, short separation
measurements of the gradient of the Casimir force between
a Au-coated sphere and a Ni-coated plate provide unique
opportunity of testing all experimental procedures for
the presence of any systematic effect which might
influence theory-experiment comparison in other
measurements.

Now we determine at what confidence level the used models
of material properties agree with the experimental
data of the experiment \cite{14}. In figure 5 the
differences of the gradients of the Casimir force (\ref{eq1})
at different experimental separations using (a) the plasma
model approach and (b) the Drude model approach are shown as the
black and gray dots, respectively. The pairs of solid lines
indicate the borders of the confidence intervals calculated
at a 67\% confidence level. As is seen in figure 5, both the
plasma and the Drude model approaches are experimentally
consistent at a 67\% confidence level.
This is not surprising taking into consideration that within
the experimental separation region both these approaches
predict almost the same gradients of the Casimir force.

Similar differences of the gradients of the Casimir force
found using (a) the plasma
model approach and (b) the Drude model approach are plotted
in figure 6, where the two solid lines show the borders of
a 10\% confidence. In both cases more than 10\% of the dots
belong to the confidence interval within any separation
subinterval. This means that in the case of Au-Ni
interacting bodies both the plasma and the Drude model
approaches agree with the measurement data at
no less than 90\% confidence level. The probability that their
common prediction is incorrect is less than 10\%.
This invalidates an assumption \cite{52} proposed for the
configuration of Au-Au test bodies that there may be some
unaccounted systematic effect (supposedly originating from
large surface patches) which superimposes on the Casimir
force and effectively brings the data in agreement with
the plasma model approach. It is true that for Au-Au
surfaces, where the two theoretical approaches predict
different gradients of the Casimir force, this is
formally possible. However, for Au-Ni test bodies any
unaccounted systematic effect  would bring the
measurement data in
disagreement with both the Drude and the plasma model
approaches.
Recently the voltage distribution due to surface patches
on the planar Au samples used in Casimir force measurements
was directly measured by means of Kelvin probe force
microscopy \cite{52a}. It was shown that the gradient of
respective additional force is more than one order of
magnitude smaller and has a stronger separation dependence
than the difference in theoretical predictions of the
plasma and Drude model approaches.

\section{Agreement with a  model of material properties
in the Casimir experiment with graphene-coated substrate}

Graphene is a two-dimensional sheet of carbon atoms having
unusual electrical, mechanical and optical properties \cite{53}.
It was shown \cite{54} that for graphene the
thermal effect in the Casimir interaction
becomes dominant at much shorter separation than
for ordinary materials, either dielectric or metallic.
This is the reason why graphene is of much promise for
experiments on measuring the Casimir force.
The material properties of graphene cannot be described by the
dielectric permittivity depending only on the frequency.
In the framework of the Dirac model \cite{53} the fundamental
description of the response of graphene on the electromagnetic
field is described by the polarization tensor \cite{55}.
The equivalent description can be formulated in terms of the
density-density correlation functions and nonlocal dielectric
permittivities depending on both the frequency and the wave
vector \cite{56}.

The first measurement of the Casimir interaction in the
configuration including a graphene sheet was performed very
recently \cite{20}. In this experiment the gradient of the
Casimir force between a Au-coated sphere of $R=54.1\,\mu$m
radius and a graphene sheet deposited on a SiO${}_2$ film
covering a Si plate has been measured. Theoretical description
of the Casimir interaction for layered structures, where some
layers are described by the frequency-dependent dielectric
permittivities and some other by the polarization tensor,
was elaborated in 2014 \cite{57}. Using this theory, the
measurement data of \cite{20} were shown to be in agreement
with the predictions of the Lifshitz theory \cite{57}, but only
the first method of theory-experiment comparison considered
in section 2 has been used. Here, we consider the differences
between theoretical and mean experimental gradients of the
Casimir force and determine more rigorously the measure of
agreement between experiment and theory.

 In figure 7 the differences between the gradients of
theoretical and mean experimental Casimir forces are shown
by dots for (a) the first and (b) the second graphene-coated
sample as functions of separation.
 The two solid lines
indicate the borders of the 67\% confidence intervals.
{}From figure 7 one can only conclude that the used
theory is consistent with the measurement data for both
samples with no indication of any quantitative measure of
agreement.

To make the comparison more informative, in figure 8 we plot
the same differences of the gradients of
the Casimir force as dots, but plot as two lines
 the borders of the 10\% confidence intervals.
{}From figure 8 one can see that there are such subintervals in
the entire separation range from 224 to 500\,nm where less than
10\% of all dots belonging to these subintervals belong to the
confidence intervals. Thus, unlike the examples, considered in
sections 2--4, for a graphene-coated substrate the favorite
theory is excluded by the data at a 10\% confidence level.

To find the measure of agreement between experiment and theory
for graphene, in figure 9 we again plot
the same differences  as dots but indicate by the two lines
 the borders of the 20\% confidence intervals.
{}From figure 9 it is seen that for both samples in any
separation subinterval more than 20\% of all dots belong to
the respective confidence intervals. This means that the
theory of the Casimir interaction with graphene-coated
substrates in not excluded by the data at a 20\% confidence
level. Equivalently, one can conclude that this theory
agrees with the data at no less than 80\% confidence
level, or that it is correct with at least 80\%
probability. If to compare with experiments using metallic
surfaces, this is a lower level of confidence. This is
explained by the lack of information about the properties of
graphene sample needed to make the theory more precise (such
as the character of impurities, the value of the gap
parameter etc. \cite{20}).

\section{Conclusions and discussion}

In the foregoing we have considered the statistical methods
of comparison between experiment and theory allowing to find
an exclusion or agreement of some models of material properties
with the data of
 experiments on measuring the Casimir interaction.
This problem is gaining in importance in the context of the
puzzle in theory-experiment comparison in the Casimir
physics discussed in section 1. In previous literature the
main attention was paid to an exclusion of one of the models
at some high confidence level whereas the measures of
agreement with the data of experimentally consistent models
were not determined.

In this paper we have shown that the probability of agreement
of some model of material properties with the data
can be found by considering
the differences between theoretical and mean measured quantities
and the confidence intervals for these differences determined at
rather low (10\% or 20\%) confidence levels.
The criterium is introduced that if more than 10\% or 20\%
differences are found inside the respective confidence intervals
one can state that the theoretical model is not excluded at
these confidence levels and, thus, agrees with the data at no less
than the complementary to unity confidence levels
equal to 90\% or 80\%, respectively.

The suggested criterium was applied to the experimental data of
four experiments on measuring the gradient of the Casimir force
by means of dynamic AFM and two theoretical approaches proposed
in the literature for metallic test bodies. It was shown that
in the experiments using Au-Au and Ni-Ni test bodies \cite{15,17,18}
the Drude model approach to calculation of the Casimir
interaction is experimentally excluded at a 95\% confidence level
over some ranges of separations.
The plasma model approach agrees with the measurement data of
these experiments at no less than 90\% confidence level.
The same analysis was performed in application to measurements of
the gradient of the Casimir force between a Au-coated sphere and
a Ni-coated plate \cite{14}. Here, the Drude and the plasma model
approaches lead to almost the same predictions which agree with
the data at a 90\% confidence level over the
entire measurement range. This result allows to exclude the role
of any unaccounted systematic effect (such as patch potentials)
in theory-experiment comparison. Finally, we have shown that the
experimental data of experiment \cite{20}
performed in the configuration
of a Au-coated sphere and a graphene-coated substrate are in
agreement with theory using the polarization tensor of graphene
at the 80\% confidence level.

The proposed method of comparison between experiment and theory
can be used in other precise experiments on measurement of the
Casimir force.

\ack{The author is grateful to G.~L.~Klimchitskaya for helpful
discussions and to U.~Mohideen for agree to use the experimental
data of our joint publications.}

\newpage
\begin{figure*}[h]
\vspace*{-3.cm}
\centerline{\hspace*{2cm}\includegraphics{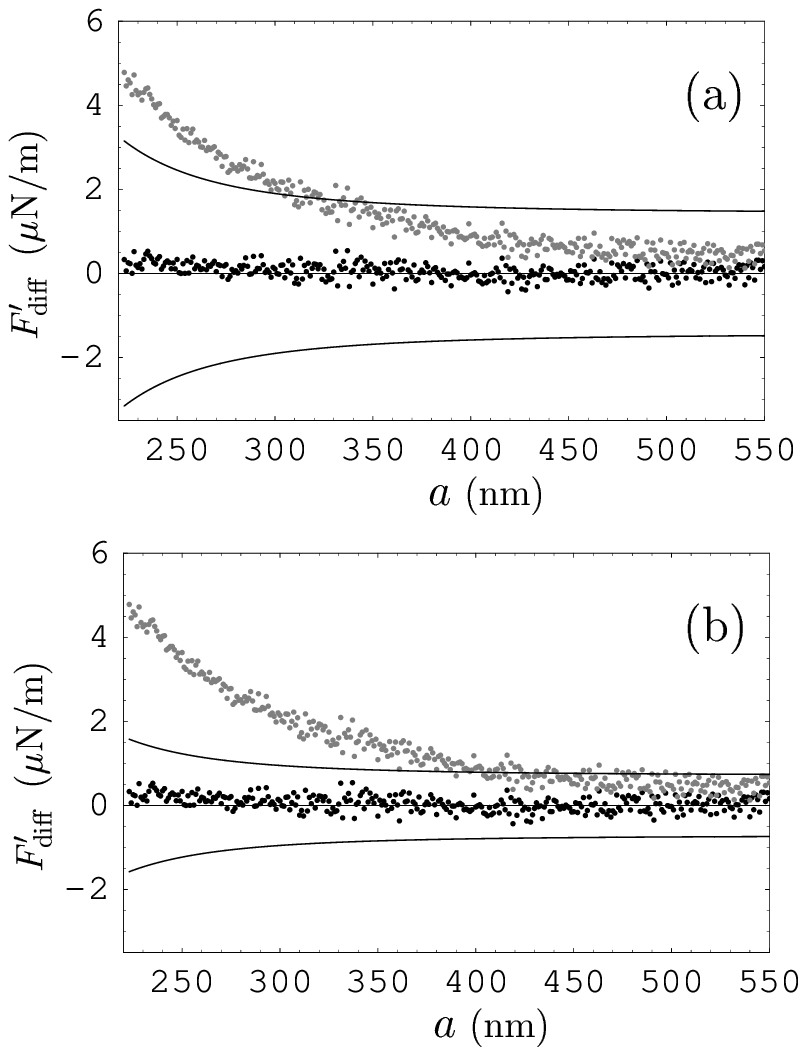}}
\vspace*{-10.cm}
\caption{
Differences between the theoretical and mean experimental
gradients of the Casimir force for Ni-Ni surfaces using the
Drude and plasma model approaches are shown by the gray and
black dots, respectively \cite{18}.
The solid lines indicate the borders of the (a) 95\% and
(b) 67\% confidence intervals.
}
\end{figure*}
\begin{figure*}[t]
\vspace*{-13.cm}
\centerline{\hspace*{4cm}\includegraphics{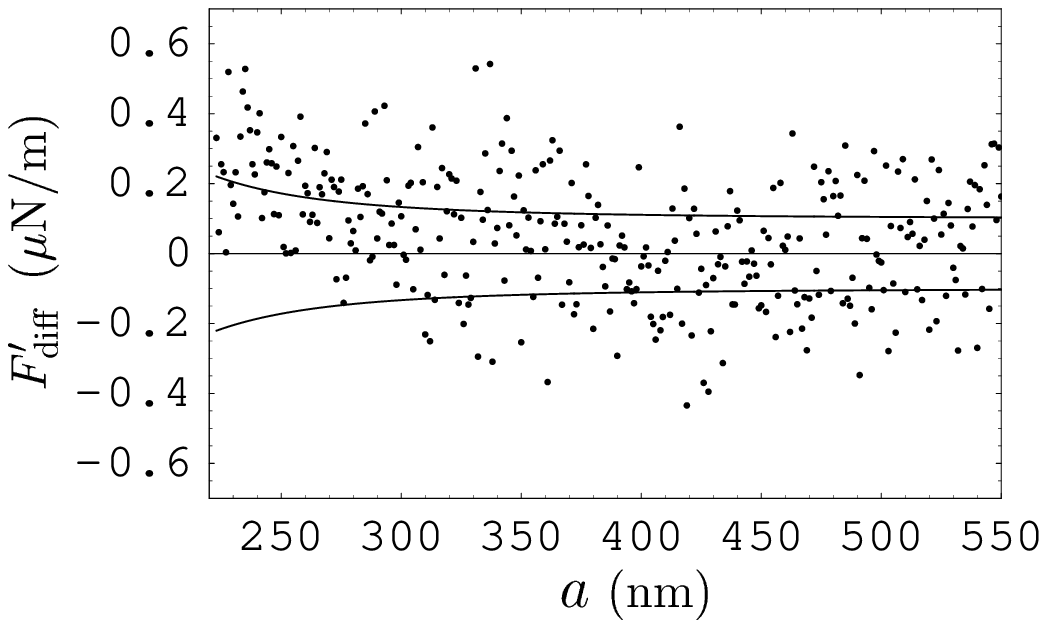}}
\vspace*{-6.cm}
\caption{
Differences between the theoretical and mean experimental
gradients of the Casimir force for Ni-Ni surfaces found using the
plasma model approach are shown by the
black dots \cite{18}.
The solid lines indicate the borders of the
10\% confidence intervals.
}
\end{figure*}
\begin{figure*}[h]
\vspace*{-3.cm}
\centerline{\hspace*{2cm}\includegraphics{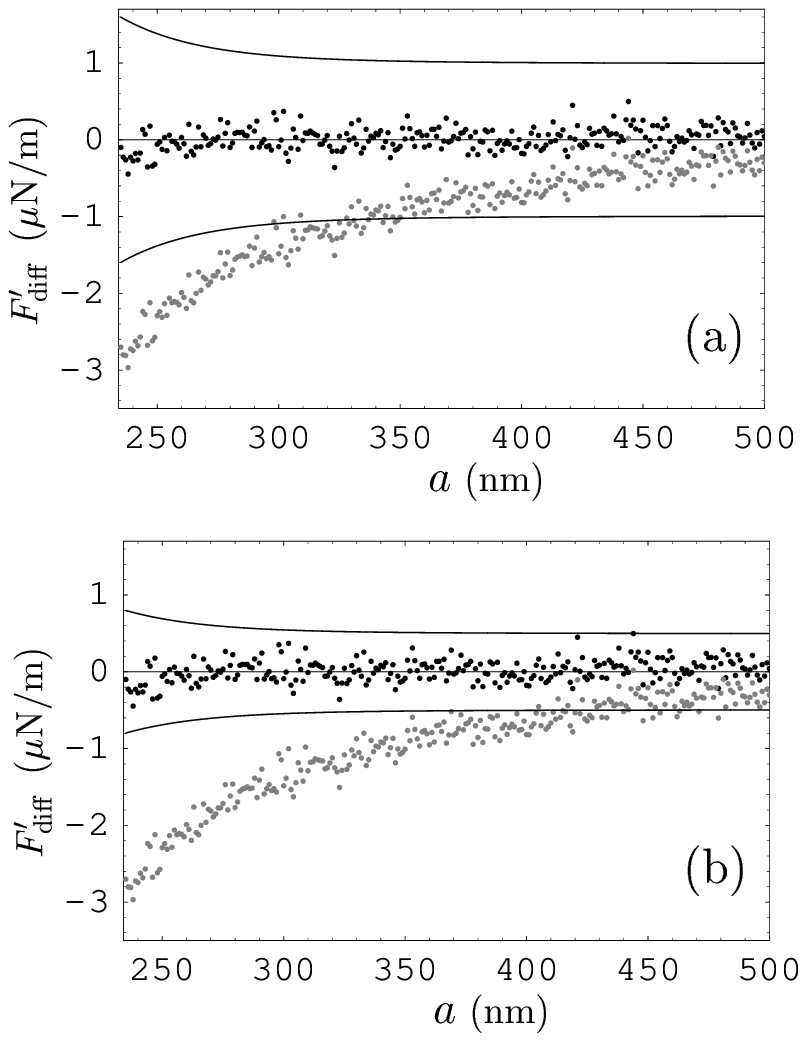}}
\vspace*{-10.cm}
\caption{
Differences between the theoretical and mean experimental
gradients of the Casimir force for Au-Au surfaces using the
Drude and plasma model approaches are shown by the gray and
black dots, respectively.
The solid lines indicate the borders of the (a) 95\% and
(b) 67\% confidence intervals.
}
\end{figure*}
\begin{figure*}[t]
\vspace*{-13.cm}
\centerline{\hspace*{4cm}\includegraphics{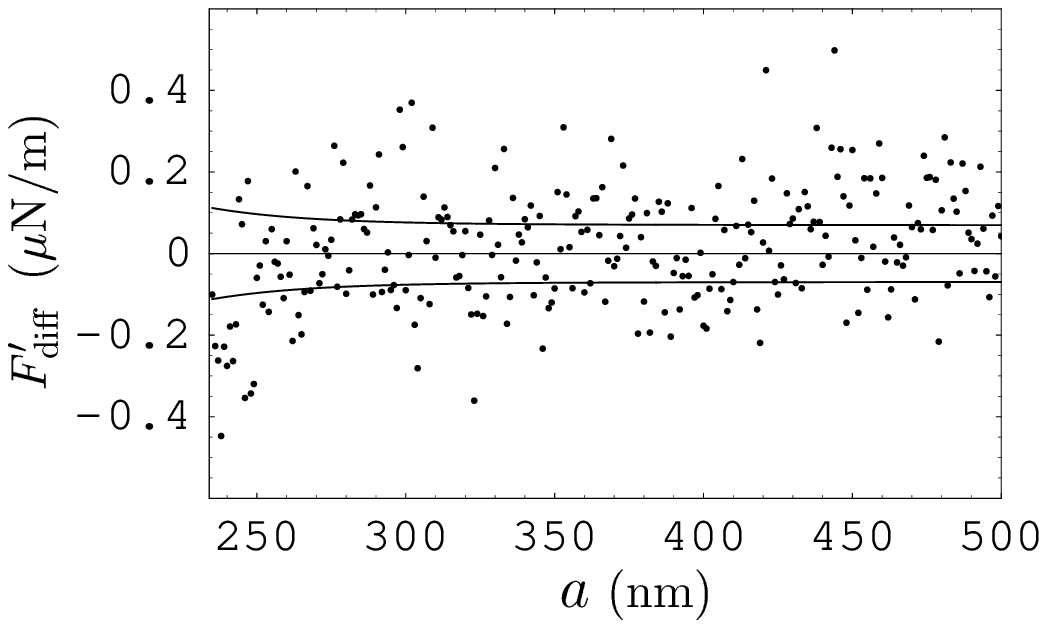}}
\vspace*{-6.cm}
\caption{
Differences between the theoretical and mean experimental
gradients of the Casimir force for Au-Au surfaces found using the
plasma model approach are shown by the
black dots.
The solid lines indicate the borders of the
10\% confidence intervals.
}
\end{figure*}
\begin{figure*}[h]
\vspace*{-3.cm}
\centerline{\hspace*{2cm}\includegraphics{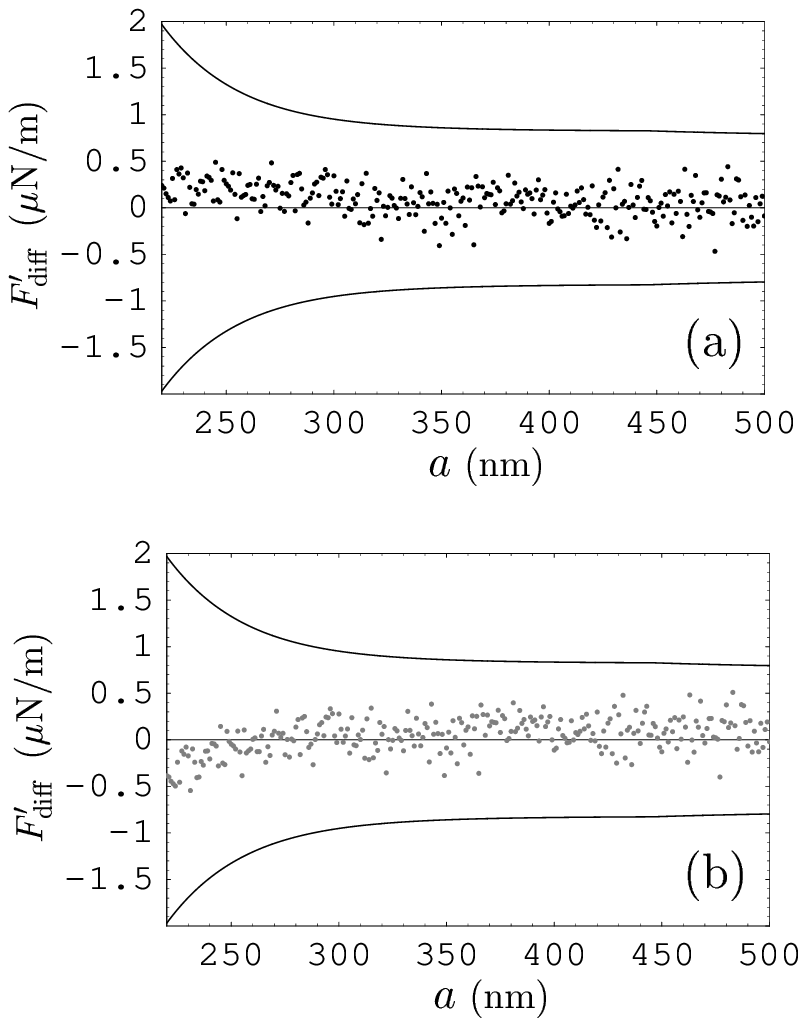}}
\vspace*{-10.cm}
\caption{
Differences between the theoretical and mean experimental
gradients of the Casimir force for Au-Ni surfaces found using
(a) the plasma and (b) the Drude model approaches are shown by the
black and gray dots, respectively.
The solid lines indicate the borders of the 67\% confidence intervals.
}
\end{figure*}
\begin{figure*}[h]
\vspace*{-3.cm}
\centerline{\hspace*{2cm}\includegraphics{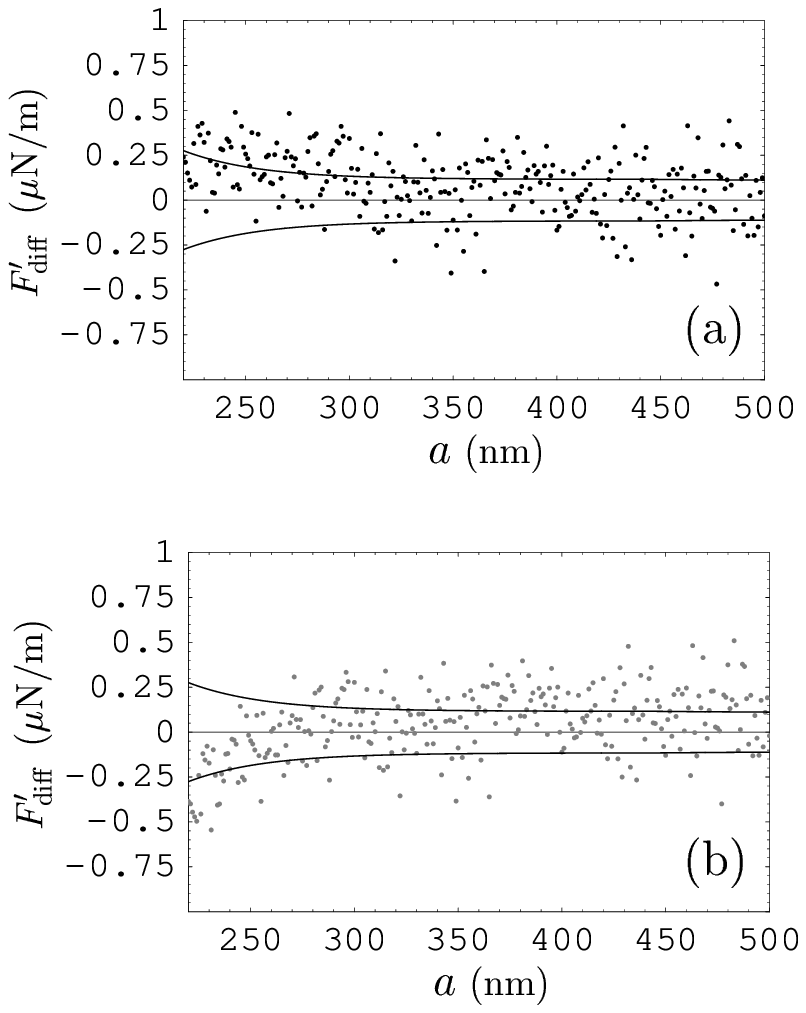}}
\vspace*{-10.cm}
\caption{
Differences between the theoretical and mean experimental
gradients of the Casimir force for Au-Ni surfaces found using
(a) the plasma and (b) the Drude model approaches are shown by the
black and gray dots, respectively.
The solid lines indicate the borders of the 10\% confidence intervals.
}
\end{figure*}
\begin{figure*}[h]
\vspace*{-3.cm}
\centerline{\hspace*{2cm}\includegraphics{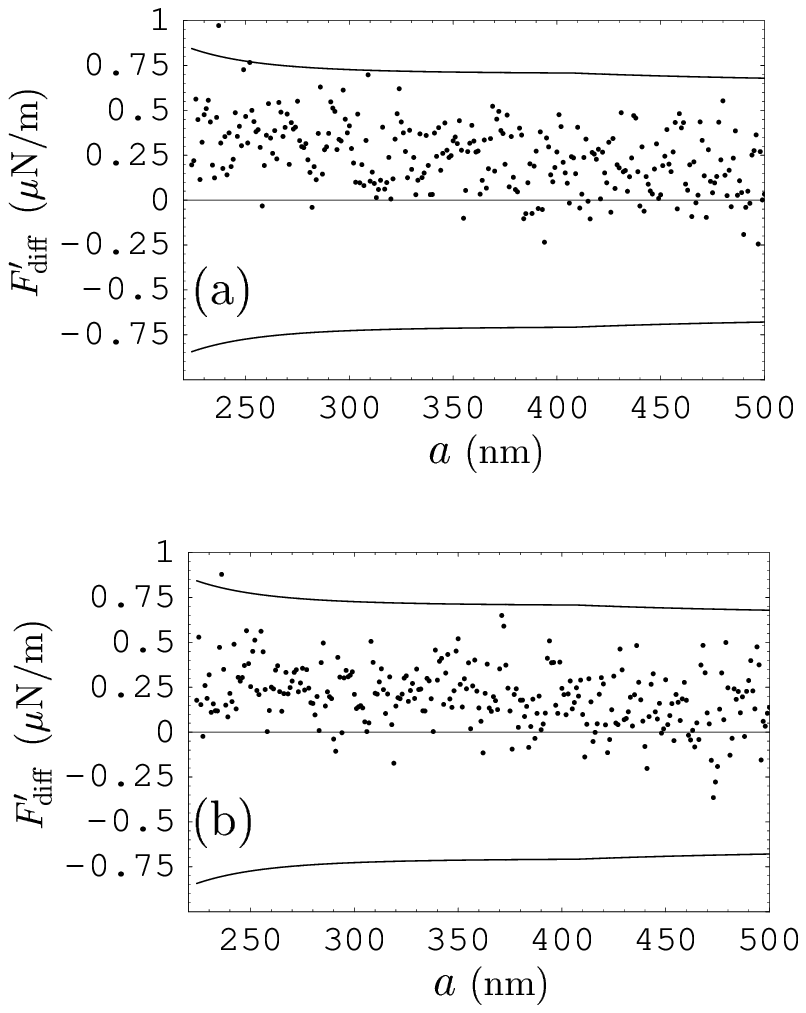}}
\vspace*{-10.cm}
\caption{
Differences between the theoretical and mean experimental
gradients of the Casimir force for the interaction of a Au-coated
sphere with a graphene-coated substrate are shown as dots for
(a) the first and (b) the second graphene-coated sample.
The solid lines indicate the borders of the 67\% confidence intervals.
}
\end{figure*}
\begin{figure*}[h]
\vspace*{-3.cm}
\centerline{\hspace*{2cm}\includegraphics{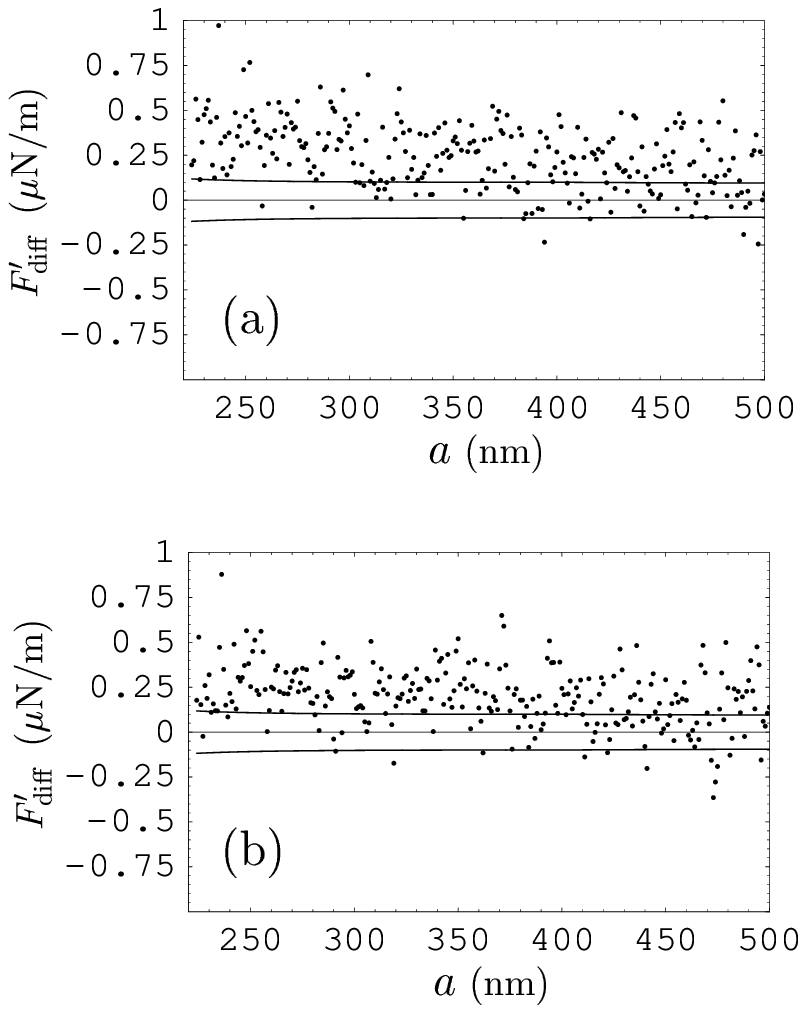}}
\vspace*{-10.cm}
\caption{
Differences between the theoretical and mean experimental
gradients of the Casimir force for the interaction of a Au-coated
sphere with a graphene-coated substrate are shown as dots for
(a) the first and (b) the second graphene-coated sample.
The solid lines indicate the borders of the 10\% confidence intervals.
}
\end{figure*}
\begin{figure*}[h]
\vspace*{-3.cm}
\centerline{\hspace*{2cm}\includegraphics{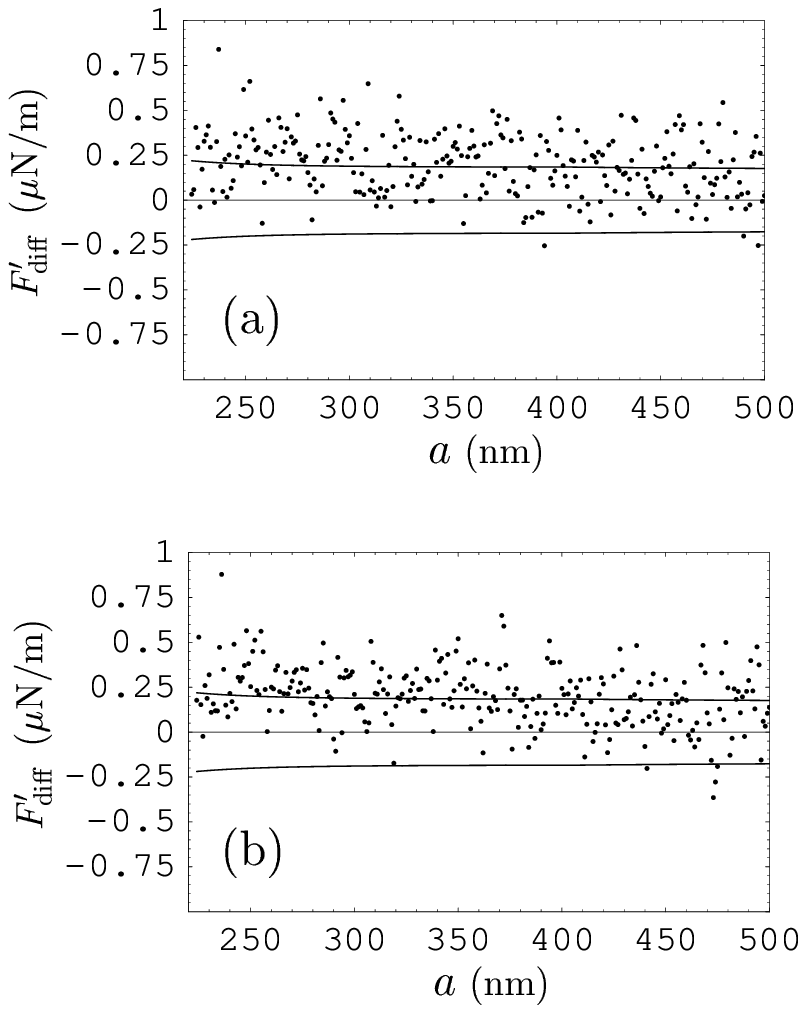}}
\vspace*{-10.cm}
\caption{
Differences between the theoretical and mean experimental
gradients of the Casimir force for the interaction of a Au-coated
sphere with a graphene-coated substrate are shown as dots for
(a) the first and (b) the second graphene-coated sample.
The solid lines indicate the borders of the 20\% confidence intervals.
}
\end{figure*}
\end{document}